\documentclass[aps,prx,reprint,amsmath,amssymb,superscriptaddress, nofootinbib, 10pt]{revtex4-2}

\usepackage[usenames,dvipsnames]{xcolor}
\usepackage{amssymb}
\usepackage{graphicx}
\usepackage{amsmath}
\usepackage{braket}
\usepackage{chemformula}
\usepackage{todonotes}
\usepackage[pdfusetitle,bookmarks=true,colorlinks,citecolor=blue,urlcolor=blue, linkcolor=blue]{hyperref}

\DeclareUnicodeCharacter{2212}{\ensuremath{-}}

\graphicspath{{Figures/}}

\newcommand{\be}{\begin{equation}}
\newcommand{\ee}{\end{equation}}

\newcommand{\meas}[1]{\mathrm{d}#1\;}

\newcommand{\eq}[1]{Eq.~(\ref{#1})}

\newcommand{\fig}[1]{Fig.\thinspace{}\ref{#1}}

\newcommand{\fc}[1]{({#1})}
\newcommand{\figc}[2]{Fig.\thinspace{}\ref{#1}\thinspace{}\fc{#2}}
\newcommand{\figcc}[3]{Fig.\thinspace{}\ref{#1}\thinspace{}\fc{#2} and \fc{#3}}
\newcommand{\figcs}[3]{Fig.\thinspace{}\ref{#1}\thinspace{}\fc{#2} - \fc{#3}}

\newcommand{\App}[1]{Appendix \ref{#1}}
\newcommand{\Apps}[2]{Appendix \ref{#1} and \ref{#2}}

\begin{document}

\newcommand{\titleinfo}{Signatures of Domain-Wall Confinement in Raman Spectroscopy of Ising Spin Chains}

\title{\titleinfo}

\newcommand{\TUM}{\affiliation{Technical University of Munich, TUM School of Natural Sciences, Physics Department, 85748 Garching, Germany}}
\newcommand{\MCQST}{\affiliation{Munich Center for Quantum Science and Technology (MCQST), Schellingstr. 4, 80799 M{\"u}nchen, Germany}}
\newcommand{\ICL}{\affiliation{Blackett Laboratory, Imperial College London, London SW7 2AZ, United Kingdom}}

\author{Stefan Birnkammer}
\email{stefan.birnkammer@tum.de}
\TUM
\MCQST

\author{Johannes Knolle}
\TUM
\MCQST
\ICL

\author{Michael Knap}
\TUM
\MCQST

\begin{abstract}
Mesonic bound states of domain walls can be stabilized in quasi one-dimensional magnetic compounds. Here, we  theoretically study the Raman light scattering response of a twisted Kitaev chain with tilted magnetic fields as a minimal model for confinement in \ch{CoNb2O6}. By both numerical matrix product states and few-domain wall variational states, we show that confinement-induced bound states directly manifest themselves as sharp peaks in the Raman response. %Near quantum criticality the Raman spectrum exhibit the famous E$_{8}$ root structure.  
Remarkably, by tuning the polarization of the incident light field, we demonstrate that the Raman response offers new insights into the intrinsic symmetry of the bound state wavefunction. 
\end{abstract}

\maketitle

\section{Introduction}
Confinement of excitations is traditionally considered as a phenomenon of high-energy physics. However, over the recent years, related effects have been discussed in condensed matter settings as well. A paradigmatic example is the formation of mesonic bound states in spin chains that exhibit a linear confinement potential between domain walls~\cite{Lake2010, Coldea2011, Wang2015, Wang2016, Bera2017, Zou2021}. A prominent candidate material to realize this effect is the quasi one-dimensional (1D) magnetic compound \ch{CoNb2O6}, often discussed as a paradigmatic example for a quantum Ising magnet in a longitudinal and transverse field~\cite{Maartense1977, Scharf1979, Heid1995, Mitsuda1994, Kobayashi1999, Kunimoto1999, Kunimoto1999b, Weitzel2000, Coldea2011, Sarvezuk2011, Kinross2014, Liang2015}.
First experimental studies of domain-wall confinement  in \ch{CoNb2O6} were achieved with neutron scattering~\cite{Coldea2011} and terahertz spectroscopy~\cite{Morris2013}. 
Further experimental and theoretical studies refined the model Hamiltonian to capture additional structure beyond the conventional Ising chain~\cite{Kjaell2011, Robinson2014,Fava2020,Laurell2022, Morris2020, Coldea2011, Cabrera2014, Amelin2020, Woodland2023, Woodland2023b}. Applying external fields, even allowed studies of confinement in the vicinity of the quantum critical point, resulting in an emergent $E_{8}$ root structure in the spectral response~\cite{Coldea2011, Amelin2020, Zou2021}.
\begin{figure*}[t!]
    \centering
    \includegraphics[width=\textwidth]{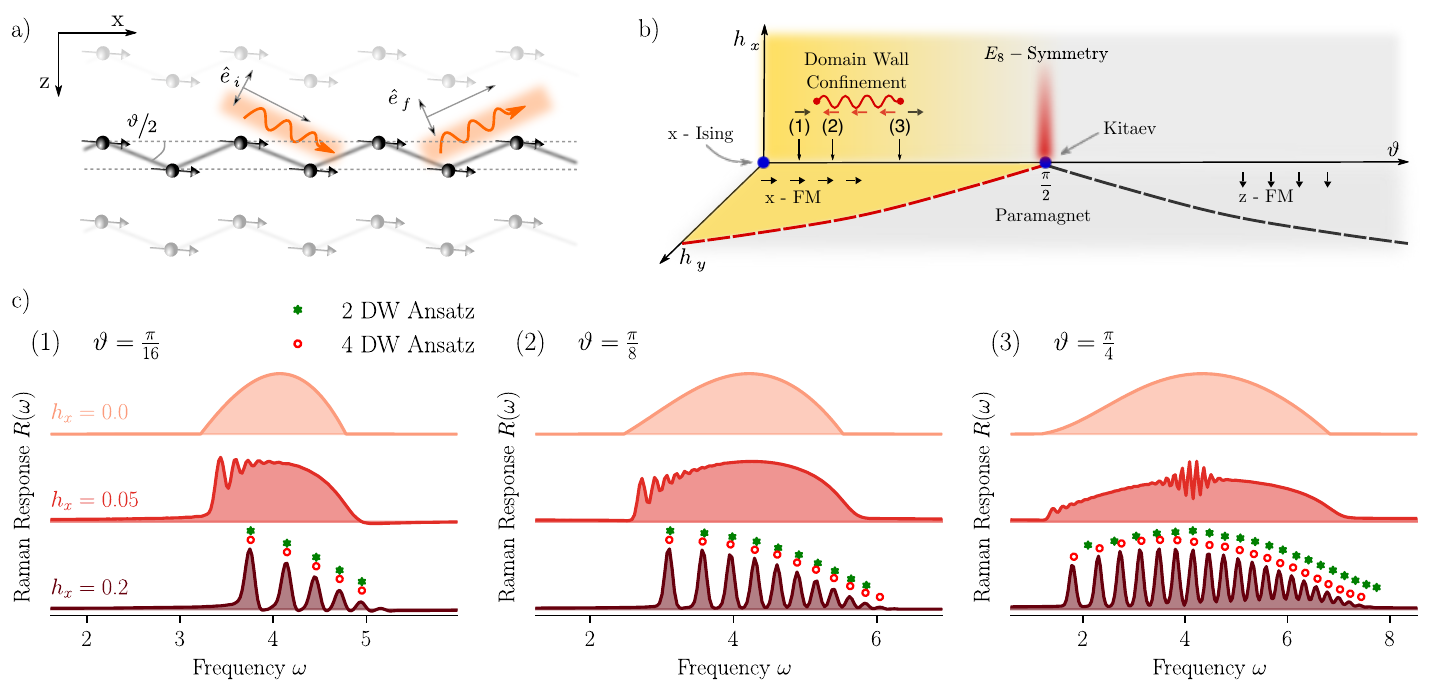}
    \caption{\textbf{Raman response of confined domain walls in a twisted Kitaev magnet.} (a) Schematic structure of a set of quasi one-dimensional chains forming \ch{CoNb2O6}, that are  characterized by a zigzag angle $\vartheta$. Raman light scattering processes of incoming photons with polarization $\hat{e}_{i}$ and outgoing photons with polarization $\hat{e}_{f}$ couple differently to even and odd bonds of the chain. (b) For a given zigzag angle $\vartheta$ and  transverse field $h_{y}$, the ground state is either a ferromagnet aligned with one of the planar direction of the chain or a paramagnet. Applying a magnetic field along the hard axis of the ferromagnet confines the low-energy domain wall (DW) excitations (yellow area for $h_x$ field). Along the Ising quantum critical line (red dashed) the spectrum is expected to possess an emergent anomalous $E_{8}$ root structure. (c) Raman response for different twist angles $\vartheta\in \{\frac{\pi}{16}, \frac{\pi}{8}, \frac{\pi}{4}\}$ and longitudinal fields $h_x = 0, 0.05$, and $0.2$ change from a two DW continuum to discrete peaks, resulting from the bound states of confined DWs. Bottom row: Deep in the ferromagnetic phase the numerical results from iMPS simulations are well captured by a few-soliton ansatz considering only 2 (green markers) or 4 (red markers) DWs.}
    \label{fig1:PhaseDiagram}
\end{figure*}

Here, we propose Raman spectroscopy as a complementary tool for studying confinement effects in solids. We consider a twisted Kitaev spin chain as a minimal model for \ch{CoNb2O6} and demonstrate that domain wall confinement manifests itself as sharp spectral peaks in the Raman response. We compute the Raman spectra numerically with Matrix Product States (MPS) as well as with few-soliton variational states. The latter correctly capture the bound state physics in the most part of the phase diagram, except near quantum criticality, where the spectrum exhibits the famous E$_8$ structure and the bound state structure is involved.  
As a key result, we show that excitations can be selectively addressed by tuning the polarization of the photons, which on the one hand provides new insights into the intrinsic structure of the bound states and on the other hand can be used to systematically study contributions to the microscopic Hamiltonian.

\section{Model}  
For our analysis we will consider a minimal model allowing us to qualitatively reproduce the key signatures observed for the magnetic compound \ch{CoNb2O6}~\cite{Morris2020}. The model captures the essential low energy physics by taking into account a zigzag geometry for the quasi one-dimensional building blocks of \ch{CoNb2O6}, see \figc{fig1:PhaseDiagram}{a}. Thereby, the spins are assumed to interact solely within a zigzag chain. The resulting model is typically referred to as the twisted Kitaev (TK) Hamiltonian 
\begin{align}
    \mathcal{H}_{\mathrm{TK}} = \sum_{j}&-\Big[ \cos^2\Big(\frac{\vartheta}{2}\Big)\sigma_{j}^{x}\sigma_{j+1}^{x} + \sin^2\Big(\frac{\vartheta}{2}\Big)\sigma_{j}^{z}\sigma_{j+1}^{z} \nonumber \\
    &+ \frac{1}{2} \sin(\vartheta) (-1)^{j}\big( \sigma_{j}^{z}\sigma_{j+1}^{x}  + \sigma_{j}^{x}\sigma_{j+1}^{z}\big) \Big].
    \label{Eq:TK-Lattice}
\end{align}
The zigzag angle $\vartheta$ interpolates between Ising ferromagnets orientated in $\hat{x}$- and $\hat{z}$-direction for $\vartheta=0$ and $\vartheta=\pi$, respectively, and a rotated version of the Kitaev chain for $\vartheta=\frac{\pi}{2}$. The weak inter-chain coupling present in actual materials is typically taken into account on a mean-field level, giving rise to an external magnetic field contribution
\be
\label{Eq:FullHamiltonian}
\mathcal{H} = \mathcal{H}_{\mathrm{TK}} - \frac{1}{2}\sum_{j} \Big[h_{x} \sigma_{j}^{x} + h_{y} \sigma_{j}^{y} + h_{z} \sigma_{j}^{z}\Big]. 
\ee
The effective magnetic fields $h_x$, $h_y$, and $h_z$ are tuneable  experimentally by external fields.
For a $\hat{y}$-polarized magnetic field, that is transverse to the ferromagnetic orders of \eq{Eq:TK-Lattice}, the model is analytically solvable by a a Jordan-Wigner transformation that maps the model to free fermions~\cite{Laurell2022, Sim2023}. The ground state phase diagram of that limit is shown in \figc{fig1:PhaseDiagram}{b} in the $(\vartheta,h_{y})$-plane. An infinitesimal magnetic field along either of the remaining two directions $\hat{x}, \hat{z}$ breaks the integrability of the model. When perturbing the ferromagnetic phases these fields can manifest themselves in two fundamentally different ways: (i) Applying an additional field transverse to the ferromagnet, increases the spatial extent of individual domain walls and modifies the dispersion law of excitations. (ii) Applying an additional field along the direction of magnetic ordering lifts the degeneracy between the ferromagnetic ground states. As a consequence, the field induces confinement between fermionic domain walls. As we will show below, both effects can be observed as characteristic signatures in the Raman response.

\section{Raman spectra}
Our analysis of the Raman response resorts to the theory by Fleury and Loudon, describing the linear response of a Mott insulator to  an external light field~\cite{Fleury1967, Fleury1968, Loudon2010}. Within this theory the Raman operator $\mathcal{R}$, creating the perturbation, contains the local spin interactions of the Hamiltonian weighted by geometric factors that reflect the polarization direction of incoming $\hat{e}_{i}$ and outgoing $\hat{e}_{f}$ photons with respect to the bond orientations $\hat{\delta}_{j}$ in the lattice
\begin{equation}\label{Eq:RamanOperatorGeneral}
    \mathcal{R} = \sum_{j} \; (\hat{\delta}_{j} \cdot \hat{e}_{i}) \;(\hat{\delta}_{j} \cdot \hat{e}_{f}) \; \mathcal{H}^{(j)},
\end{equation}
where $\mathcal{H}^{(j)}$ denote the local bond contributions to the Hamiltonian $\mathcal{H}_{\mathrm{TK}}$.
For the zigzag lattice geometry this yields two different projectors $P_{e}(\hat{e}_{i}, \hat{e}_{f}; \vartheta)$ and $P_{o}(\hat{e}_{i}, \hat{e}_{f}; \vartheta)$  for even and odd bonds, respectively. Consequently, the Raman operator then reads
\begin{equation}
    \mathcal{R} 
    =  \sum_{j} \Big[P_{e}(\hat{e}_{i}, \hat{e}_{f}; \vartheta)\;\mathcal{H}^{(2j)} + P_{o}(\hat{e}_{i}, \hat{e}_{f}; \vartheta)\;\mathcal{H}^{(2j +1)} \Big].
    \label{Eq:RamanOperatorFull}
\end{equation}
We will consider photon scattering processes in-plane with the lattice vectors of our chain. The projectors $P_{e/o}$ are then geometric factors that depend on the angle of the incoming $\theta_i$ and outgoing $\theta_f$ photons and the zigzag angle $\vartheta$; see \App{sec:RamanOperator}.
%Moreover, in the first part of our work, we will fix the polarization of both the in- and outgoing photons as follows. We set the ingoing photon polarization $\hat{e}_{i}$ to point opposite to the $\hat{x}$-direction and the outgoing photon polarizations half-way between the $\hat{x}$ and $\hat{z}$, for details see \supp  (SM). With these considerations, the photon projectors only depend on the zigzag angle $\theta$ and are given by 
% \begin{align}
% \label{Eq:ProjectorEven}
%     P_{e}(\vartheta) &= \frac{1}{\sqrt{2}}\Big[\cos^2\Big(\frac{\vartheta}{2}\Big) - \frac{\sin\big(\vartheta)}{2} \Big] \\
%     \label{Eq:ProjectorOdd}
%     P_{o}(\vartheta) &= \frac{1}{\sqrt{2}}\Big[\cos^2\Big(\frac{\vartheta}{2}\Big) + \frac{\sin\big(\vartheta)}{2} \Big].
% \end{align}
% More general polarization directions will be considered later in the work. Remarkably, we show that the polarization dependence provides insights into  the structure of the domain wall bound states. 

\noindent The Raman spectrum is obtained as follows
\be
    R(\omega) = \int \meas{t} e^{i(\omega + E_{0})t} \bra{0} \mathcal{R} e^{-i\mathcal{H}t} \mathcal{R} \ket{0},
\ee
where $\ket{0}$ is the ground state and $E_{0}$ denotes its energy. We emphasize that the Raman response couples uniformly to the system and thereby only probes excitations with zero total momentum. This is because Raman spectroscopy operates in the visible regime, leading to a momentum transfer that is much smaller than a typical reciprocal lattice momentum. %Hence, in order to fulfill energy and momentum conservation, light can only create pairs of excitations with opposite momentum. 
We use the translational invariance of the excited state $\mathcal{R}\ket{0}$ to efficiently compute the response numerically using iMPS methods~\cite{Sim2022, Sim2023}, for details see \App{ssec:iMPSDetails}. 
\begin{figure}
    \centering
    \includegraphics[width=\columnwidth]{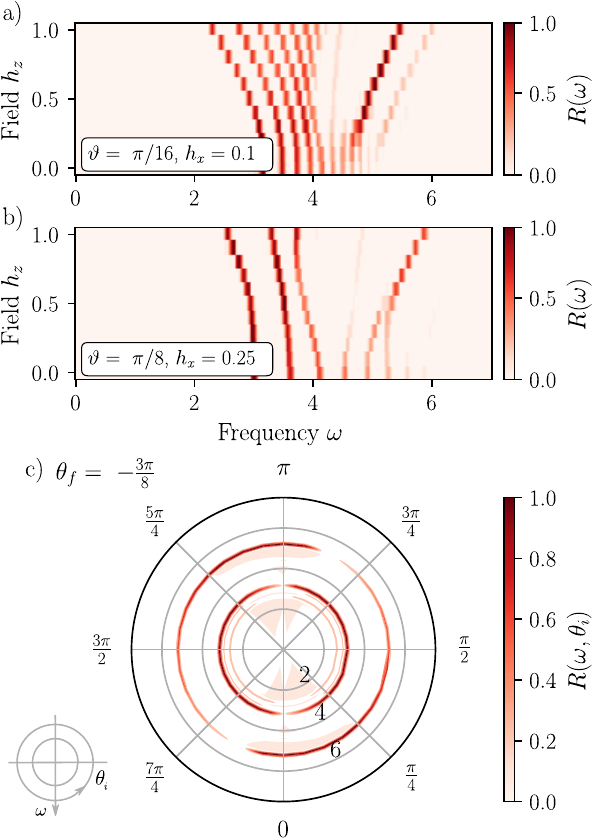}
    \caption{\textbf{Analyzing the structure of bound states.} 
    Transverse field $h_{z}$ dependence of bound state energies for (a) $\vartheta=\frac{\pi}{16},\; h_{x}=0.1$, $h_{y}=0.5$ and (b) $\vartheta=\frac{\pi}{8},\; h_{x}=0.25$, $h_{y}=0.5$. The $h_{z}$ field splits the peak structure into three sets, whose energy either decreases, stays approximately constant, or increases with increasing $h_{z}$. (c) Polar plot of the Raman spectrum as a function of the polarization of incident photons $\theta_{i}$ and fixed outgoing polarization $\theta_{f}=-\frac{3\pi}{8}$. Tuning the polarization of the scattered light, suppresses for certain angles the contribution of some of the bound states. The  results are obtained for $\vartheta=\frac{\pi}{8}, (h_x, h_y, h_z)=(0.1, 0.0, 2\sin(\vartheta))$.}
    \label{fig3:EnrichedDynamics}
\end{figure}\\
We now focus on a fixed polarization of the light field. Specifically, we set the ingoing photon polarization $\hat{e}_{i}$ to be the $-\hat{x}$-direction and the outgoing photon polarization $45^\circ$ between the $\hat{e}_{i}$ and $\hat{z}$ directions. With these considerations, the photon projectors only depend on the zigzag angle $\vartheta$ and are given by $P_{e} = \frac{1}{\sqrt{2}}[\cos^2({\vartheta}/{2}) - {\sin(\vartheta)}/{2} ]$ and $P_{o} = \frac{1}{\sqrt{2}}[\cos^2({\vartheta}/{2}) + {\sin(\vartheta)}/{2} ]$.
% \begin{align}
% \label{Eq:ProjectorEven}
%     P_{e}(\vartheta) &= \frac{1}{\sqrt{2}}\Big[\cos^2\Big(\frac{\vartheta}{2}\Big) - \frac{\sin\big(\vartheta)}{2} \Big] \\
%     \label{Eq:ProjectorOdd}
%     P_{o}(\vartheta) &= \frac{1}{\sqrt{2}}\Big[\cos^2\Big(\frac{\vartheta}{2}\Big) + \frac{\sin\big(\vartheta)}{2} \Big].
% \end{align}
% More general polarization directions will be considered later in the work. Remarkably, we show that the polarization dependence provides insights into  the structure of the domain wall bound states. 
The Raman spectra for vanishing transverse magnetic fields $h_{y}$ and $h_{z}$, different values of $h_{x}$, and zigzag angles $\vartheta\in \{\frac{\pi}{16}, \frac{\pi}{8}, \frac{\pi}{4}\}$  are shown in \figc{fig1:PhaseDiagram}{c}. The angle $\vartheta$ is chosen such that the ground state in the absence of the $h_x$-field is in the $\hat{x}$-ferromagnetic phase. Our results illustrate how the Raman spectrum crosses over from a continuum in the absence of a longitudinal magnetic field $h_{x}=0$ to a discrete sequence of bound state peaks for increasing $h_{x}$. The results for $h_x=0$ are computed analytically and results for finite $h_x$ are obtained numerically with iMPS (see \Apps{sec:IntegrabliltyRegime}{ssec:iMPSDetails}).\\
%The difference in frequency between two consecutive peaks, thereby, also shows a direct dependence on $h_{x}$. To confirm the physical picture of the discrete peak structure arising from the formation bound state of two solitonic domain walls we benchmark the results against exact diagonalization (ED) results allowing only a certain number of solitons. 
%To obtain an intuition of the observed discrete peak structure we compare our results to a simple trial wavefunction describing a small number of solitonic domain wall excitations. 
Next we benchmark our numerically exact tensor network results against the response using a trial wavefunction in a restricted Hilbert space of only a few solitonic domain wall excitations.
To construct the basis for this ansatz we start with the ground state configuration of all spins  aligned in $\hat{x}$-direction and systematically add pairs of solitons, i.e., pairs of domain walls, to the state until we reach a chosen maximal number of domain walls, see \App{ssec:FKEDDetails}. The projected Hamiltonian restricted to this few soliton subspace is expected to reproduce the low-energy response of the system when the energy cost associated to domain walls is large compared to their kinetic energy. This assumption is valid deep in the ferromagnetic regime and in particular far away from the Ising critical line separating the ordered from the paramagnetic phase.\\
The peak structure of the Raman spectrum  obtained within this few-soliton ansatz captures the exact iMPS results remarkable well for strong longitudinal fields; see \figc{fig1:PhaseDiagram}{c}. Already the minimum of two domain walls (green stars) capture quantitatively the peak positions for $\vartheta=\frac{\pi}{16}$. However, as the excitation energy for domain walls decreases with increasing $\vartheta$,  deviations between the iMPS and the two-soliton ansatz are observed. Including further domain walls significantly improves the accuracy of the predicted peaks (red circles: four domain walls). Hence, the Raman response captures the bound state formation of domain wall excitations and thus serves as a direct probe of domain wall confinement. When approaching the critical point $(\vartheta=\frac{\pi}{2})$ more and more solitons need to be included to reproduce the bound state spectrum. This suggests that the entire spectrum of multi-soliton excitations becomes strongly dressed in the vicinity of the critical point.

\section{Characterizing the structure of bound states with Raman scattering}

So far we have investigated the consequences of a confining magnetic field component $h_{x}$ on the $\hat{x}$-polarized FM state. %Next we analyze the polarization dependence of the Raman response. To this end, we consider additional transverse fields.
Applying in addition a transverse field $h_{z}$ splits the bound state spectrum into three sets of resonances, as has been previously detected in experiments~\cite{Morris2020, Woodland2023}. This splitting does not occur for a twisted Kitaev Hamiltonian with a longitudinal field alone. We illustrate the splitting for a zigzag angle $\vartheta=\frac{\pi}{16}$ and longitudinal field $h_{x}=0.1$ as well as for $\vartheta=\frac{\pi}{8}$ and $h_{x}=0.25$ as a function of $h_z$ in \figcc{fig3:EnrichedDynamics}{a}{b}. The splitting results from different internal symmetries of the bound state wave functions~\cite{Woodland2023}. In the $\hat{x}$-polarized FM phase for a large transverse $h_z$ field the effective Hamiltonian dimerizes. The dimerized bound states factorize into three sets of states, that shift toward lower energies, remain constant, and shift toward higher energies as $h_z$ is increased. These states are then even, invariant, and odd with respect to bond inversion $U$, see \App{sec:PolarizationRaman}. %The flow of the bound state energies with $h_z$ field is also seen in the numerical results of \figcc{fig3:EnrichedDynamics}{a}{b}; for more details see \supp. 
We now determine the Raman response when tuning the polarization of the incoming photons $\theta_i$; see  \figc{fig3:EnrichedDynamics}{c}, where we used a fixed polarization of outgoing photons $\theta_{f}=-\frac{3\pi}{8}$, and $\vartheta=\frac{\pi}{8}, (h_x, h_y, h_z)=(0.1, 0., 2 \sin \vartheta)$. We find that for certain angles of the polarization, the Raman signal associated to specific bound states can be completely suppressed. 
\begin{figure}
    \centering
    \includegraphics[width=\columnwidth]{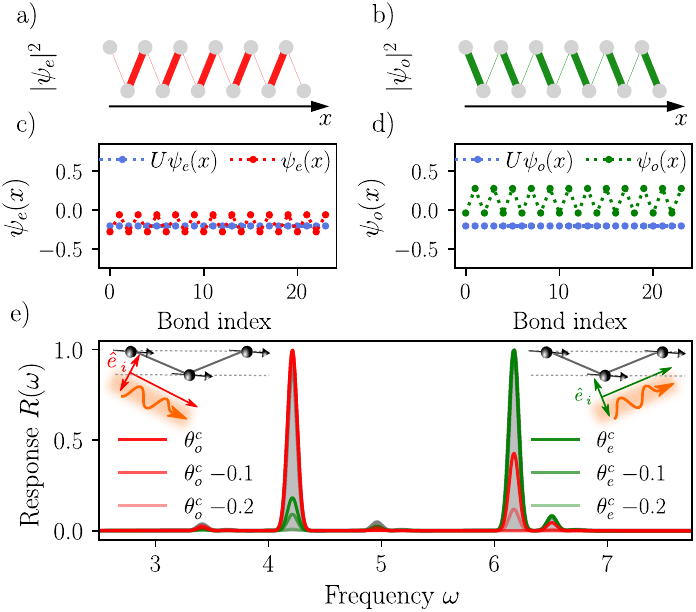}
    \caption{\textbf{Selection criteria for localized bound states.} (a) - (b) Real-space soliton  distributions $\vert\psi_{e}\vert^2$,  $\vert\psi_{o}\vert^2$ for even and odd bound states, respectively. For even bound states, solitons are predominantly located on even bonds and vice versa for odd states. (c) - (d) Under bond inversion $U$ the phase of odd bound states exhibits characteristic sign change between neighboring sites, whereas even bound states preserve the phase. (e) Demonstration of the symmetry selection criteria. The spectral response of even bound states (peak at $\omega \approx 4.2$) diminishes when the incoming photon polarization $\theta_{e}^{c}$ is chosen perpendicular to even lattice bonds (green curves). By contrast, the response of odd bound states (peak at $\omega \approx 6.2$) vanishes for a polarization perpendicular to odd bonds (red curves).}
    \label{Fig3:PolarizationResolved}
\end{figure}
In order to understand this behavior, we explore the internal structure of the bound states using a two soliton ansatz, which qualitatively reproduces the exact numerical results. From this ansatz we find the eigentstates ${\psi}_{e/o}$, that are even and odd under bond inversion $U$, respectively, to inherit the dimerized structure of the Hamiltonian, \figcc{Fig3:PolarizationResolved}{a}{b}. The bound states are predominantly localized on even (odd) bonds for even (odd) eigenstates. We furthermore illustrate the bond inversion operation $U$ on the wave functions, which unveils a characteristic staggered phase between neighboring bonds for odd bound states and a uniform phase for even bound states;  \figcc{Fig3:PolarizationResolved}{c}{d}. \\
Due to the zigzag geometry of the Ising chain, light couples differently to even and odd bonds. When tuning the polarization of the light field, the coupling to the bond-centered bound states is thereby tuned. Light with a polarization perpendicular to the bond cannot couple to a wave function localized on the respective bond and the Raman response of this bound state vanishes. We demonstrate this by fixing the photon polarization to $\theta_{e}^{c}$ chosen such that even bonds are projected out, which reduces coupling to even eigenstates (green curves in \figc{Fig3:PolarizationResolved}{e}). Choosing the photon polarization to be perpendicular to odd bonds  $\theta_{o}^{c}$ projects out the response attributed to odd bound states (red curves). These observations are robust to weak deviations of the polarization from the optimal angles as shown in \figc{Fig3:PolarizationResolved}{e}.

\section{Discussion \& Outlook}
We have demonstrated that Raman light scattering couples to   bound states of domain walls in confined Ising spin chains and, hence, can be used as a complementary tool to neutron scattering and THz spectroscopy. Remarkably, by tuning the polarization of the scattered light field the internal symmetries of the excitations are probed. This additional information can be used to refine the structure of the effective microscopic model. The polarization dependence of the Raman spectrum can also be used to detect terms beyond the Ising field theory near quantum critical points. At criticality, the Raman spectrum resolves the emergent low-energy E$_8$ root structure of the Ising field theory as we demonstrate in \App{sec:E8Structure}. The eigenstates of the Ising field theory are agnostic to the polarization dependence of the scattered light. However, the spectral weight of bound states that arise from contributions beyond the Ising field theory can be tuned by the polarization. This enables one to systematically study contributions to the microscopic Hamiltonian beyond the low-energy Ising theory.

Our work has focused on the twisted Kitaev chain in a tilted field, which is a minimal description of \ch{CoNb2O6}.  
In this system the splitting of the peaks in the Raman spectrum is of the order of $0.1~\text{meV}$~\cite{Coldea2011, Morris2013}, which can be resolved with available Raman setups~\cite{Hackl2022, Jost2022, He2022, Versteeg2019}. However, our results go beyond this specific material. For example, recent studies of the quasi one-dimensional antiferromagnet \ch{BaCo2V2O8} revealed  similar phenomena of mesonic domain-wall bound states. Due to the larger antiferromagnetic interactions, however, the peak splitting of bound states is even of the order of $0.5~\text{meV}$~\cite{Faure2018, Zhang2020, Zou2021, Wang2023}. This renders \ch{BaCo2V2O8} a complimentary candidate material for observing bound states of domain walls in inelastic light scattering experiments with even more favorable energy scales.
In future work, it will be interesting to study how the selection rules resulting from the polarization of the light fields manifest themselves in other magnetic compounds with confinement.

\section*{Acknowledgements}
We thank G. B. Sim, J. Hausschild and F. Pollmann for interesting discussions about the model and details on the numerical methods. We thank R. Hackl and P. v. Loosdrecht for comments on possible experimental implementations. We acknowledge support from the Deutsche Forschungsgemeinschaft (DFG, German Research Foundation) under Germany’s Excellence Strategy--EXC--2111--390814868, TRR 360 – 492547816 and DFG grants No. KN1254/1-2, KN1254/2-1, the European Research Council (ERC) under the European Union’s Horizon 2020 research and innovation programme (grant agreement No. 851161), as well as the Munich Quantum Valley, which is supported by the Bavarian state government with funds from the Hightech Agenda Bayern Plus.

\section*{Data and materials availability}
Data, data analysis, and simulation codes are available upon reasonable request on Zenodo~\cite{zenodo}.

\begin{appendix}

\section{The Raman Operator}
\label{sec:RamanOperator}

In this section we derive the Raman operator for a generic choice of photon polarizations relative to the sample. Within the Fleury-Loudon theory of Raman spectroscopy, the Raman operator $\mathcal{R}$ is constructed from interactions between two neighboring spins. These are weighted by the projection of the individual photon polarizations onto the bond connecting the spins. Following the notation of the main text, the Raman operator reads
\begin{align}
\label{Eq:RamanOperatorFL}
    \mathcal{R} = \sum_{j=1}^{L} (\hat{\delta}_{j} \cdot \hat{e}_{i}) \;\mathcal{H}^{(j)}\; (\hat{\delta}_{j} \cdot \hat{e}_{f}) \equiv \sum_{j=1}^{L} P_{j}(\hat{e}_{i}, \hat{e}_{f}; \vartheta) \;\mathcal{H}^{(j)}
\end{align}
Here, we denote the unit vectors describing the bond orientation in our zigzag chain by $\hat{\delta_{j}}$. A suitable parameterization of $\hat{\delta_{j}}$ as well as the polarization vectors is given by
\begin{align}
\hat{\delta_{j}} = \begin{pmatrix}
 \cos\Big(\frac{\vartheta}{2}\Big) \\
 0 \\
 (-1)^{j} \sin\Big(\frac{\vartheta}{2}\Big)
\end{pmatrix}  &  &
\hat{e}_{i/f} = \begin{pmatrix}
 \sin(\theta_{i/f}) \cos(\phi_{i/f}) \\
 \sin(\theta_{i/f}) \sin(\phi_{i/f}) \\
 \cos(\theta_{i/f})
\end{pmatrix}.
\end{align}
With this convention the projectors $P_{j}(\hat{e}_{i}, \hat{e}_{f}; \vartheta)$ included in \eq{Eq:RamanOperatorFL} assume two distinct values for even and odd bonds, respectively
\begin{align}
\label{Eq:RamanProjectorsGeneral1}
    P_{e}(\hat{e}_{i}, \hat{e}_{f}) &= \Big[\cos\Big(\frac{\vartheta}{2}\Big)\sin(\theta_{i}) \cos(\phi_{i})+ \sin\Big(\frac{\vartheta}{2}\Big)\cos(\theta_{i})\Big] \nonumber\\
    &\times\Big[\cos\Big(\frac{\vartheta}{2}\Big)\sin(\theta_{f})\cos(\phi_{f}) + \sin\Big(\frac{\vartheta}{2}\Big)\cos(\theta_{f})\Big] 
  \nonumber \\
    P_{o}(\hat{e}_{i}, \hat{e}_{f}) &= \Big[\cos\Big(\frac{\vartheta}{2}\Big)\sin(\theta_{i})\cos(\phi_{i}) - \sin\Big(\frac{\vartheta}{2}\Big)\cos(\theta_{i})\Big] \nonumber\\
    &\times\Big[\cos\Big(\frac{\vartheta}{2}\Big)\sin(\theta_{f})\cos(\phi_{f}) - \sin\Big(\frac{\vartheta}{2}\Big)\cos(\theta_{f})\Big].
\end{align}
Thus the Raman operator takes the simple functional form denoted in \eq{Eq:RamanOperatorFull} of the main text. Considering only photon scattering processes in the plane spanned by $\hat{\delta_{1}}$ and $\hat{\delta_{2}}$, moreover, fixes $\phi_{i} = \phi_{f} = 0$. In general, we find that the Raman operator inherits the two-site unit cell of the Hamiltonian with local contributions $\mathcal{R}^{(j)}$ acting on three spins
\be
\label{Eq:RamanDensity}
\mathcal{R} = \sum_{j=1}^{L/2} \Big[ P_{e}(\hat{e}_{i}, \hat{e}_{f}) \mathcal{H}^{(2j)} + P_{o}(\hat{e}_{i}, \hat{e}_{f}) \mathcal{H}^{(2j+1)} \Big] \equiv  \sum_{j=1}^{L/2} \mathcal{R}^{(j)}.
\ee

\section{Numerical Methods}
\subsection{Raman Response from iMPS Simulations}
\label{ssec:iMPSDetails}

\noindent For finite systems with confinement, single-domain wall excitations can be stabilized in the vicinity of the boundaries. To avoid these artifacts we develop infinite system size matrix product states (iMPS). We use the TeNPy library as a basic framework for our simulations~\cite{Hauschild2018}. Our approach is conceptually similar to earlier works~\cite{Sim2022, Sim2023} and consists of the following steps:
\begin{enumerate}
    \item Compute the ground state $\ket{0}$ with energy $E_{0}$ of the system for given parameters of Hamiltonian~(1) using the iDMRG algorithm.
    \item Blow-up the tensor to $L$.
    \item Compute the set of states $I = \{\mathcal{R}^{(j)}\ket{0}\}_{1\le j < L}$ obtained from applying the local Raman operator $\mathcal{R}^{(j)}$ defined in \eq{Eq:RamanDensity} to site $j$ of the iMPS.
    \item Perform a time evolution of the specific state $\mathcal{R}^{(L/2)}\ket{0}$ using the iTEBD algorithm.
    \item Calculate the gauge-fixed overlap between the time evolved state and all states of $I$ by calculating the dominant eigenvalue of the transfer matrix within the spatial window of size $L$, i.e., compute $C^{(j)}(t) \equiv e^{iE_{0}t} \bra{0} \mathcal{R}^{(j)} e^{-i\mathcal{H}t}\mathcal{R}^{(L/2)}\ket{0}$ for all values of $j$~\cite{Kjaell2011, Sim2022, Sim2023}.
    \item Making use of the translational invariance manifest in the iMPS formalism compute $R(t) = L \sum_{j=1}^{L} C^{(j)}(t)$.
\end{enumerate}
\begin{figure*}
    \centering
    \includegraphics[width=\textwidth]{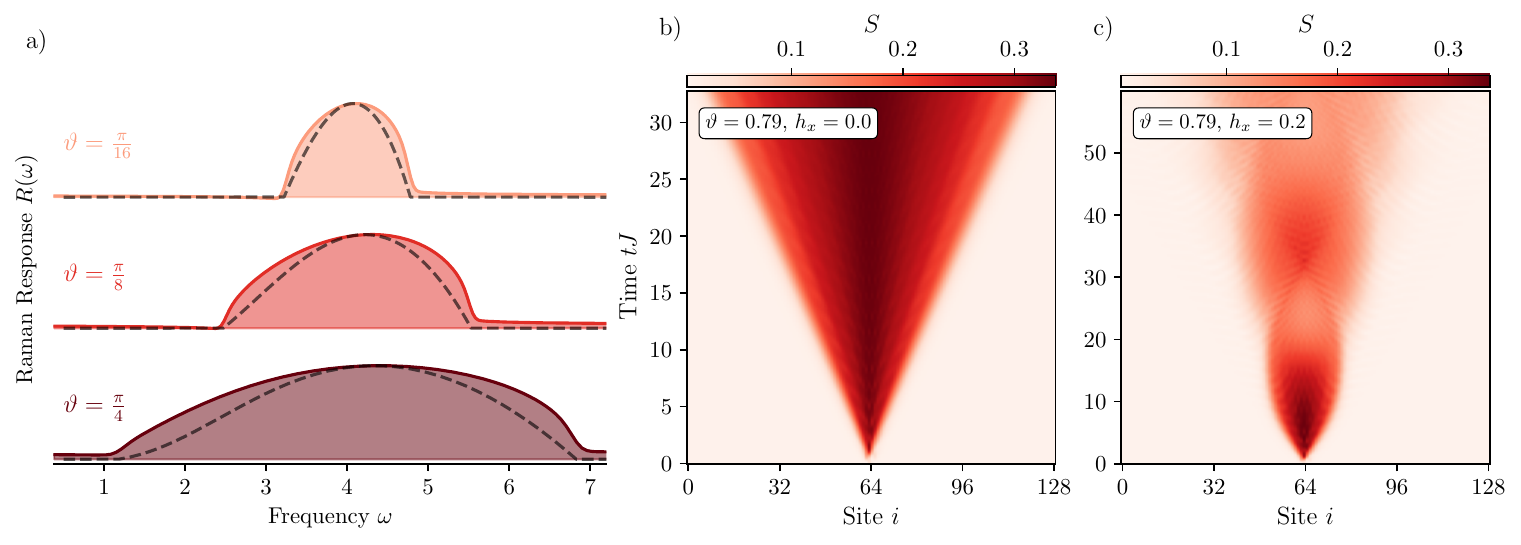}
    \caption{\textbf{Raman response obtained from iMPS in the analytically solvable limit.} (a) We benchmark the results obtained from the iMPS approach (colored areas) with analytical predictions (gray dashed lines) for different values of $\vartheta\in \{\frac{\pi}{16}, \frac{\pi}{8}, \frac{\pi}{4}\}$ and $h_{x}\!=\!h_{y}\!=\!h_{z}\!=\!0$. Within our simulations, using a unit cell of $L=128$ sites, we find the energy range of excitations as well as the frequency for the largest Raman response to be well reproduced by iMPS. Qualitative differences, however, are evident in the functional form of the spectral response. This discrepancy is a result of the short simulation times we can reach due to fast entanglement growth in the integrable regime. (b) The lightcone spreading of entanglement after applying the local Raman density in the center of the unit cell in the absence of a longitudinal field limits the maximal simulation times to $tJ\sim 30$ for the unit cell size $L=128$. (c) In presence of a longitudinal field entanglement growth is drastically reduced due to the confining potential between pairs of excited fermions. As a consequence, our simulations yield accurate results up to at least double as long times of $tJ\sim60$ as demonstrated for the example of $\vartheta=\frac{\pi}{4}, h_{x}=0.2$ and $h_{y}=h_{z}=0$.}
    \label{fig:Benchmark_iMPS_Analytics}
\end{figure*}
In the last step we have used that the tensors in the iMPS are translational invariant. This is what ultimately allows us to efficiently compute the Raman response probing the zero momentum response of the system. We obtain converged results in the regime of confinement for bond dimensions of $\chi=200$ and all data shown in this manuscript is obtained for $\chi=400$. In fact the most challenging regime for the iMPS method is the free fermion case in absence of a confining field, as the excitations created by the Raman operator can travel with a high velocity through the systems. In that limit, we indeed find  differences between analytical predictions and numerical results, as shown in \figc{fig:Benchmark_iMPS_Analytics}{a}. Although, the total energy regime for excitations are captured well, the iMPS results cannot reproduce the precise form of the response due to the comparatively short accessible simulations times. Ballistic spreading of entanglement limits the times up to which we can simulate with a given unit cell of length $L$. Once a longitudinal field is applied, the spectral response stays confined to a finite region of space for long times, reducing the effective growth of entanglement. The behavior of the entanglement growth without and with an applied longitudinal field is shown in \figcc{fig:Benchmark_iMPS_Analytics}{b}{c}. 

\subsection{Few Domain-Wall Ansatz}
\label{ssec:FKEDDetails}

To obtain physical insights into the numerical results obtained from tensor network methods, we have also implemented a few domain wall ansatz. This approach projects the Hilbert space to a maximal domain wall number and then uses  exact diagonalization on a finite system with periodic boundary conditions. Concretely, we construct a basis for the Hilbert space starting from the ferromagnetic configurations and successively adding pairs of domain walls to the states. It is worth pointing out that this construction explicitly avoids configurations of odd soliton number, which for the confined phases would lead to irrelevant high-energetic excitations. With this procedure the full low energy Hilbert space can be spanned. In practice we stop the construction with $N$ domain walls and project the model into this subspace ending up with a Hamiltonian $\mathcal{H}_{N}$~\cite{Birnkammer2022}.\\
\begin{figure*}
    \centering
    \includegraphics[width=1.0\textwidth]{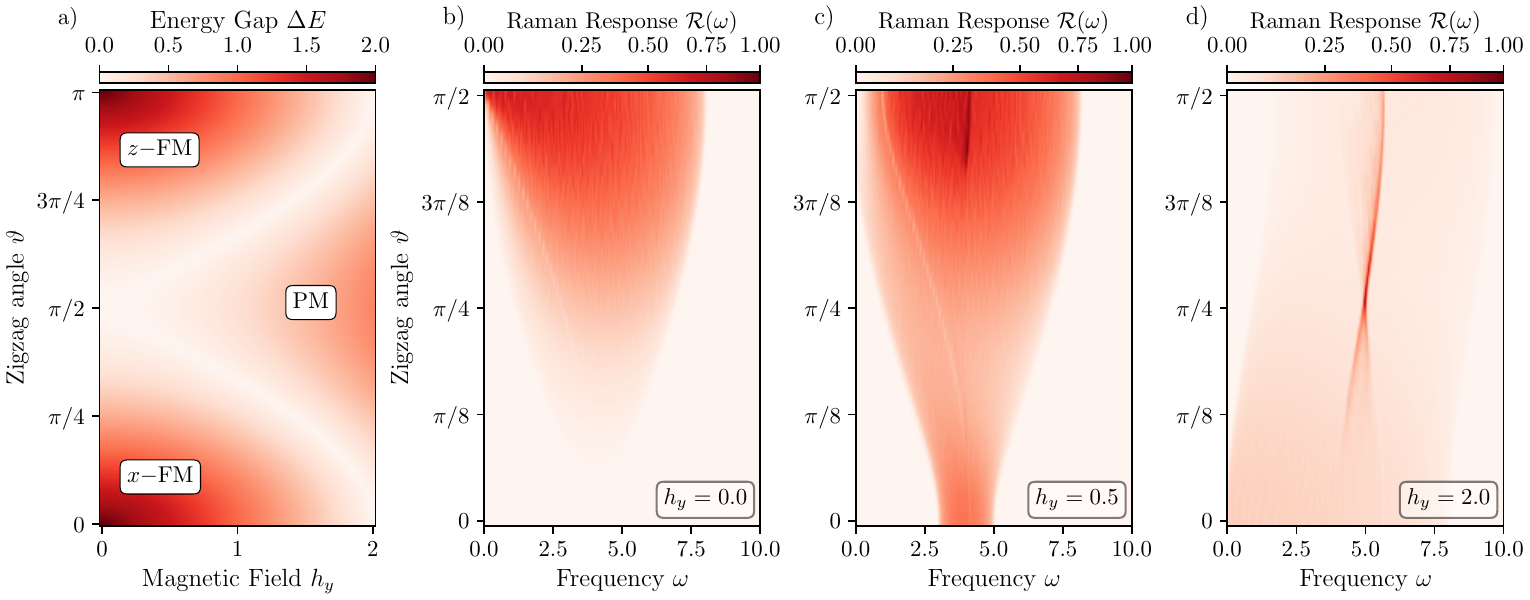}
    \caption{\textbf{Analytical results for phase diagram and Raman response in the integrable limit.} (a) We show the analytical result for the energy gap $\Delta E$ for the integrable regime $h_{x}=h_{z}=0$ as a function of zigzag angle $\vartheta$ and transverse field $h_{y}$. There is an Ising critical line separating the two ferromagnetic phases from a magnetically disordered paramagnetic phase. (b) - (d) Analytical predictions for the Raman response $R(\omega)$ for different values of $h_{y}$ are shown as a function of frequency $\omega$ and zigzag angle $\vartheta$. While in the FM phase the Raman response is dominated by the two-domain wall continuum, the spectrum of the PM is dominated by scattering processes involving the flat band of the Kitaev model, associated with a divergent density of states.}
    \label{fig:RamanSpectra}
\end{figure*}
This approach has two main advantages; first and foremost it enables us to simulate exactly comparatively large systems. The dominant contribution to the subspace dimension results from the sector of largest domain wall number $N$, which scales with system size as $\propto\binom{L}{N}$. Comparing this to the exponential scaling of the full Hilbert space ($\propto 2^{L}$) we can thus simulate much larger values of $L$ with comparable computational resources. Secondly, stepwise increasing the number of solitons taken into account in our simulation can give us additional information about the nature of low-energy quasiparticles in our system. This is also discussed in the main text in \figc{fig1:PhaseDiagram}{c}. There, we find that for small values of $\vartheta=\frac{\pi}{16}$ already 2 domain walls accurately reproduce the peak structure of the iMPS calculations, indicating that bound states of only two solitons describe the low-energy response of the model. For larger values of $\vartheta=\frac{\pi}{4}$  contributions from higher domain wall number sectors are required. The true bound states are thus dressed excitations. When approaching quantum criticality the structure of the confined bound states becomes increasingly complex as discussed in \App{sec:ExcitationsCriticality}. \\

\section{Raman Response in the Integrable Regime}
\label{sec:IntegrabliltyRegime}

\subsection{Spectrum of the Exactly Solvable Model}
\label{ssec:SpectrumModel}

In the presence of the transverse $h_y$ field only, our system is integrable. We now analytically compute the Raman response for this regime. On the one hand, we will use these results to quantitatively determine the phase boundaries sketched in \figc{fig1:PhaseDiagram}{b}. On the other hand, we can use them to benchmark the accuracy of our numerical results obtained with iMPS. We emphasize already now that due to the absence of confinement and the fast speed of excitations, this integrable case is actually the most difficult case for iMPS.
For the analytical solution we make use of a Jordan-Wigner transformation and map the spin model to dual fermions. The latter can be solved by a Bogoliubov transformation, which directly allows us to compute the Raman response.
We define the Jordan-Wigner mapping in y-direction to end up with a theory of well defined fermion parity
\begin{align}
\sigma_{i}^{y} &= 2 c_{i}^{\dagger}c_{i} - 1  \nonumber \\
\sigma_{i}^{x} &= \prod_{j<i} (1- 2 c_{j}^{\dagger}c_{j}) (c_{i} + c_{i}^{\dagger}) \nonumber \\
\sigma_{i}^{z} &= -i \prod_{j<i} (1- 2 c_{j}^{\dagger}c_{j}) (c_{i} - c_{i}^{\dagger}).
\end{align}
Inserting this transformation in the Hamiltonian (1) for $h_{x}=h_{z}=0$ we find 
\begin{widetext}
\vspace{-\baselineskip}
\begin{align}
\mathcal{H}(\vartheta) &= -\sum_{j=1}^{L} \Big[c_{j}^{\dagger}c_{j+1} + \Big(\cos(\vartheta) + i (-1)^{j} \sin(\vartheta)\Big) c_{j}^{\dagger}c_{j+1}^\dagger  + \text{h.c.} \Big] - \frac{h_{y}}{2} \sum_{j=1}^{L} \Big[2 c_{j}^{\dagger}c_{j} - 1\Big] \nonumber \\
 &= - \sum_{j=1}^{L/2} \Big[b_{j}^{\dagger}a_{j} + a_{j}^{\dagger}b_{j+1} + e^{i\vartheta}b_{j}^{\dagger}a_{j}^\dagger + e^{-i\vartheta}a_{j}^{\dagger}b_{j+1}^\dagger  + \text{h.c.} \Big] - \frac{h_{y}}{2} \sum_{j=1}^{L/2} \Big[2 a_{j}^{\dagger}a_{j} + 2 b_{j}^{\dagger}b_{j} - 2\Big].
\end{align}
\end{widetext}
In the last line we have introduced operators for the unit cell on odd ($a_{j} = c_{2j+1}$) and even ($b_{j}=c_{2j}$) sites. Using the translational invariance of the two-site unit cell, the Fourier transformation is given by  
\begin{align}
a_k &= \sqrt{\frac{2}{L}} \sum_{j=1}^{L/2} e^{ikj} a_{j} & b_k &= \sqrt{\frac{2}{L}} \sum_{j=1}^{L/2} e^{ikj} b_{j}.
\end{align}
Then the momentum space representation of $\mathcal{H}$ reads
\begin{widetext}
\vspace{-\baselineskip}
\begin{align}
    \mathcal{H}(\vartheta) &= -\sum_{k} \Big[e^{i\vartheta}b_{k}^{\dagger} a_{-k}^{\dagger} + b_{k}^{\dagger} a_{k} + e^{-i(\vartheta - k)} a_{-k}^{\dagger} b_{k}^{\dagger} +  e^{-i k} a_{k}^{\dagger} b_{k}+ \text{h.c.} \Big] - \frac{h_{y}}{2}\sum_{j} \Big[2 a_{k}^{\dagger}a_{k}  + 2 b_{k}^{\dagger}b_{k}\Big] - \frac{h_{y}}{2} L \nonumber \\
    &\equiv  - \sum_{k} \Big[ A(k)b_{k}^{\dagger}a_{k} + A(k)^{*}a_{k}^{\dagger}b_{k} + B(k, \vartheta)b_{k}^{\dagger}a_{-k}^{\dagger} + B(k, \vartheta)^{*}a_{-k}b_{k} \Big] + \mathcal{C},
\end{align}
\end{widetext}
where $A(k)= 1 + e^{ik}$ and $B(k, \vartheta)= e^{i\vartheta} - e^{i(k - \vartheta)} = \cos(\vartheta) (1-e^{ik}) + i \sin(\vartheta) (1+e^{ik}) \equiv B_{1}(k, \vartheta) + B_{2}(k, \vartheta)$ depend on the momentum and the zigzag angle $\vartheta$. We also introduced a variable accounting for the overall constant factor $\mathcal{C}= -\frac{K}{2}h_{y}L$. Additionally we find a parity symmetry reflected in $A(k) = A(-k)^{*}$.\\
We can symmetrize the expression to bring it to conventional Bogoliubov-deGennes form using the spinor notation $\psi_{k}^{\dagger} = (a_{k}^{\dagger}, a_{-k}, b_{k}^{\dagger}, b_{-k})$
\be
\label{Eq:BdG-Hamiltonian}
\mathcal{H}_{BdG}(\vartheta)= -\frac{1}{2}\sum_{k} \psi_{k}^{\dagger}\;\Gamma(k; \vartheta)\;\psi_{k} 
\ee
with the geometric coupling matrix 
\be
\Gamma(k, \vartheta) = \begin{pmatrix}
 h_{y} & 0 & A(k)^{*} & -B(-k, \vartheta) \\
 0 & -h_{y} & B(k, \vartheta)^{*} & -A(k)^{*}\\
 A(k) & B(k, \vartheta) & h_{y} & 0 \\
 -B(-k, \vartheta)^{*} & -A(k) & 0 & -h_{y}
\end{pmatrix}.
\ee
After diagonalizing the Hamiltonian~(\ref{Eq:BdG-Hamiltonian}) using the eigenmodes of $\Gamma(k, \vartheta)$ we find
\be
\label{Eq:DiagonalExpressionBdG}
\mathcal{H}_{BdG}(\vartheta) = \sum_{k} \Big[\gamma_{k} \Big(\alpha_{k}^\dagger \alpha_{k} - \frac{1}{2}\Big) + \rho_{k}\Big(\beta_{k}^{\dagger}\beta_{k} -\frac{1}{2}\Big) \Big], 
\ee
where $\alpha_{k}, \beta_{k}$ are the operators contained in the diagonalizing Nambu spinor $\phi_{k}^{\dagger} = (\alpha_{k}^{\dagger}, \alpha_{-k}, \beta_{k}^{\dagger}, \beta_{-k})$ defined via the unitary transformation $\Lambda(k, \vartheta)$ diagonalizing $\Gamma(k, \vartheta)$
\begin{align}
\Lambda(k, \vartheta)^{-1} \Gamma(k, \vartheta) \Lambda(k, \vartheta) &= \text{diag}(\gamma_{k}, -\gamma_{-k}, \rho_{k}, -\rho_{-k})  \nonumber \\ \phi_{k}^{\dagger} &= \psi_{k}^{\dagger} \Lambda(k, \vartheta).
\end{align}
The eigenenergies thereby relate to the functions $A(k)$ and $B(k, \vartheta)$ as
\begin{align}
\label{Eq:Dispersion-TK}
\gamma_{k} &= \sqrt{h_{y}^2 + \xi_{k} - \sqrt{4\lambda_{k} h_{y}^{2} + \xi_{k}^2 - \eta_{k}^2}}  \nonumber \\
\rho_{k} &= \sqrt{h_{y}^2  + \xi_{k} + \sqrt{4\lambda_{k} h_{y}^{2} + \xi_{k}^2 - \eta_{k}^2}}
\end{align}
\begin{figure*}[t!]
    \centering
    \includegraphics[width=\textwidth]{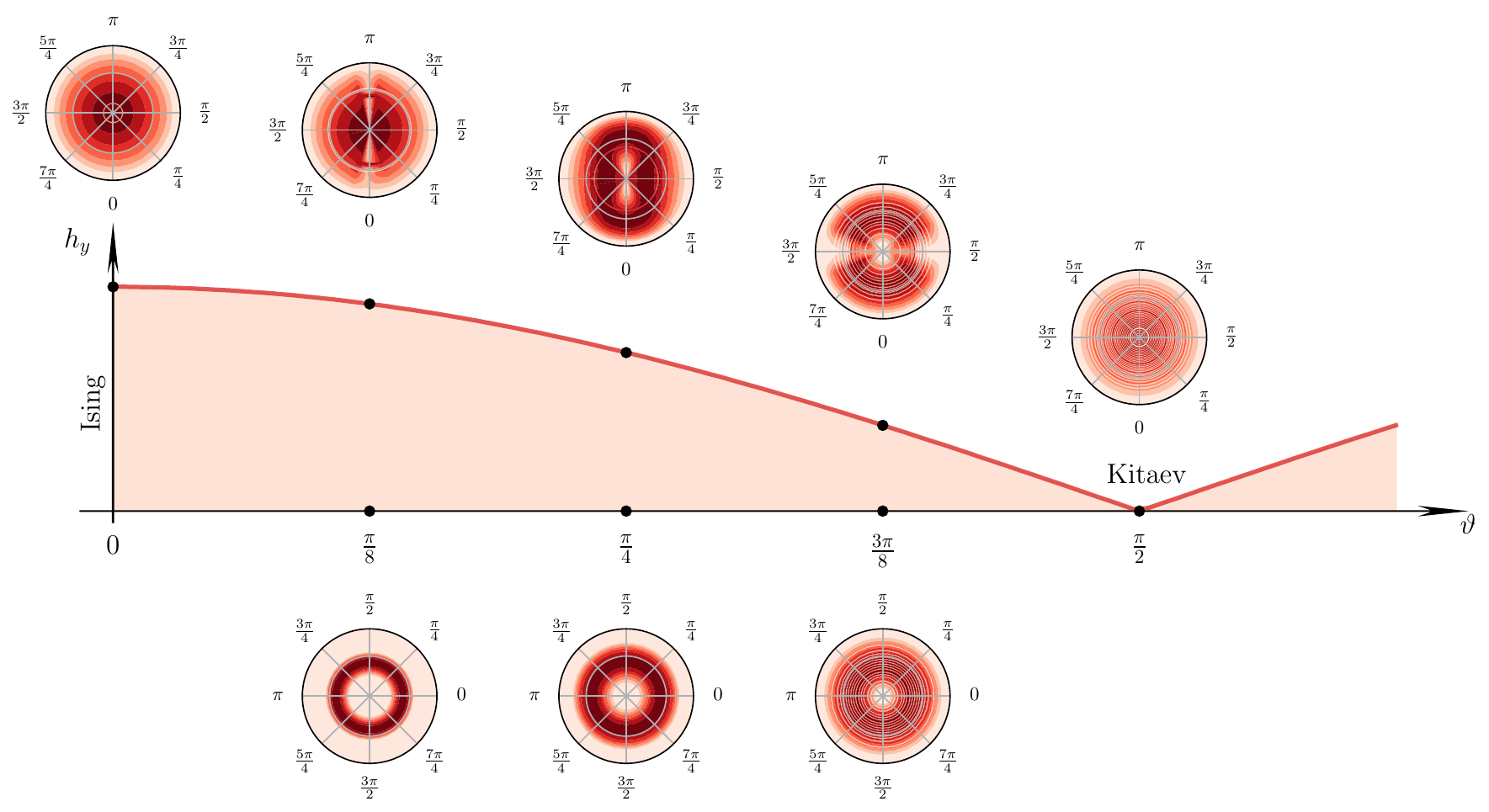}
    \caption{\textbf{Polarization dependent Raman spectra for the integrable model.} We show angular resolved Raman spectra for different regions in the phase diagram for the analytically tractable $(h_{y},\vartheta)$-plane ($h_{x}=h_{z}=0$). All results are obtained for fixed polarization of ingoing photons aligned with the $\hat{z}$-direction. The polarization of outgoing photon is tuned in the $(\hat{x}, \hat{z})$-plane, i.e. varying $\theta_{f}$ while keeping $\phi_{f}=0$. We find the rotational symmetry of the response to be broken if we consider both finite zigzag angles, giving rise to two distinct photon projectors $P_{e}$ and $P_{o}$, as well as finite transverse fields causing a staggering of hopping matrix elements in the chain. Rotational symmetry of the response is thus generically broken at the quantum critical point, but gets restored at the extreme points $\vartheta=0$ (Ising) and $\vartheta=\frac{\pi}{2}$ (Kitaev). This structure gets inherited for situations including finite confinement fields $h_{x}$, where $E_{8}$ peaks resulting from Ising field theory show no particular angular dependence, but excitations associated with zigzag angle of the chain inherit angular dependence.}
    \label{Fig8:PolarizationDependenceUnconfined}
\end{figure*}
with  
\begin{align}
\xi_{k}&=\vert A(k)\vert^2 +\vert B_{1}(k)\vert^2 +\vert B_{2}(k)\vert^2  \nonumber\\
\eta_{k}&=\vert A(k)^2 - B_{1}(k)^2 + B_{2}(k)^2\vert  \nonumber \\
\lambda_{k} &= \vert A(k)\vert^2 = 2 - 2\cos(k) .
\end{align}
The dispersion denoted in \eqref{Eq:Dispersion-TK} reduces in the limiting case of $\vartheta=0$ to the well-known Ising dispersion for a transverse magnetic field $h_{y}$
\be
\gamma_k \rvert_{\vartheta = 0} = 2 \sqrt{1 + \Big(\frac{h_{y}}{2}\Big)^2 \pm h_{y}\cos\Big( \frac{k}{2} \Big)}.
\ee
For the pure Kitaev case ($\vartheta=\pi/2$) we find 
\begin{align}
\big(\gamma_k \rvert_{\vartheta = \frac{\pi}{2}}\big)^2 &= 8 \cos^2\Big(\frac{k}{2}\Big) + h_{y}^2  \nonumber  \\
&\pm 4 \Big\vert \cos\Big(\frac{k}{2}\Big)\Big\vert \sqrt{h_{y}^2 + 4 \cos^2\Big(\frac{k}{2} \Big)}
\end{align}
including the zero mode for the Kitaev model for the minus branch in the case of $h_y = 0$.
The quantitative phase diagram resulting from the dispersion~(\ref{Eq:Dispersion-TK}) is shown in \figc{fig:RamanSpectra}{a}.

\subsection{Raman Response of the Dual Fermionic Theory}
\label{ssec:RamanResponseAnalytical}

To evaluate the Raman spectra, the Raman operator has to be expressed in the diagonal basis of the Hamiltonian \eq{Eq:DiagonalExpressionBdG}. First, we rewrite the expression for the Raman operator of \eq{Eq:RamanDensity} in the fermionic basis 
\begin{align}
\mathcal{R}(\vartheta) = \sum_{j}^{L/2} \Big[ &P_{o}(\hat{e}_{i}, \hat{e}_{f}; \vartheta) \big( b_{j}^{\dagger}a_{j}^{\dagger} e^{i\vartheta} + b_{j}^{\dagger}a_{j}+\text{h.c.} \big)  \nonumber \\
+ &P_{e}(\hat{e}_{i}, \hat{e}_{f}; \vartheta) \big( a_{j}^{\dagger}b_{j+1}^{\dagger} e^{-i\vartheta} + a_{j}^{\dagger}b_{j+1} +\text{h.c.} \big)  \Big].
\end{align}
Transforming the expression into momentum space we end up with
\begin{align}
\mathcal{R}(\vartheta) &= \sum_{k} \Big[\big( P_{o}(\hat{e}_{i}, \hat{e}_{f}; \vartheta) +  P_{e}(\hat{e}_{i}, \hat{e}_{f}; \vartheta) e^{ik} \big) b_{k}^{\dagger}a_{k} \nonumber \\ 
+ \big( &P_{o}(\hat{e}_{i}, \hat{e}_{f}; \vartheta)e^{i\vartheta} -  P_{e}(\hat{e}_{i}, \hat{e}_{f}; \vartheta)e^{i(k-\vartheta)}\big) b_{k}^{\dagger}a_{-k}^{\dagger} + \text{h.c.}\Big].
\end{align}
This is similar to the result we obtained when diagonalizing the Hamiltonnian with the minor difference that we have to include projectors for the photon directions into the definitions of $A(k)$ and $B(k, \vartheta)$
\begin{align}
\tilde{A}(k) &=  P_{o}(\hat{e}_{i}, \hat{e}_{f}; \vartheta) +  P_{e}(\hat{e}_{i}, \hat{e}_{f}; \vartheta) e^{ik} \nonumber \\
\tilde{B}(k, \vartheta) &= P_{o}(\hat{e}_{i}, \hat{e}_{f}; \vartheta)e^{i\vartheta} -  P_{e}(\hat{e}_{i}, \hat{e}_{f}; \vartheta)e^{i(k-\vartheta)}
\end{align}
In this notation, the Raman operator takes the form 
\be
\mathcal{R}(\vartheta) = \frac{1}{2}\sum_{k} \psi^{\dagger}_{k} \;\tilde{\Gamma}(k; \vartheta, \hat{e}_{i}, \hat{e}_{f}) \;\psi_{k}
\ee
and $\tilde{\Gamma}(k; \vartheta, \hat{e}_{i}, \hat{e}_{f})$ having fundamentally the same form as the $\Gamma(k; \vartheta)$ coupling matrix of the Hamiltonian with new coefficients $A(k)\rightarrow \tilde{A}(k), B(k, \vartheta)\rightarrow \tilde{B}(k, \vartheta)$
\be
\tilde{\Gamma}(k; \vartheta) = 
\begin{pmatrix}
 0 & 0 & \tilde{A}(k)^{*} & -\tilde{B}(-k, \vartheta) \\
 0 & 0 & \tilde{B}(k, \vartheta)^{*}& -\tilde{A}(k)^{*}\\
 \tilde{A}(k) & \tilde{B}(k, \vartheta) & 0 & 0 \\
-\tilde{B}(-k, \vartheta)^{*} & -\tilde{A}(k) & 0
\end{pmatrix}.
\ee
In the diagonal basis of $\mathcal{H}$ this changes according to the basis change $\Lambda(k;\vartheta)$ resulting from the Bogoliubov transformation
\be
\label{eq:RamanDiagonalBasis}
\mathcal{R}(\vartheta) = \sum_{k} \phi^{\dagger}_{k} \; \tilde{D}(k; \vartheta, \hat{e}_{i}, \hat{e}_{f})\; \phi_{k}
\ee
with $\tilde{D}(k; \vartheta, \hat{e}_{i}, \hat{e}_{f}) = \Lambda(k;\vartheta)^{\dagger}  \tilde{\Gamma}(k; \vartheta, \hat{e}_{i}, \hat{e}_{f}) \Lambda(k;\vartheta)$.\\
\eq{eq:RamanDiagonalBasis} allows us to evaluate the Raman response for a generic eigenstate of the exactly solvable model. Here we are, however, primarily interested in the ground state response of the system and compute 
\begin{align}
R(\omega) &= \int\meas{t} e^{i(\omega + E_{0})t} \bra{0} \mathcal{R} e^{-i\mathcal{H}t} \mathcal{R} \ket{0} \nonumber\\
&= \int\meas{t} e^{i(\omega + E_{0})t} \sum_{k, q} \tilde{D}_{\alpha \beta}(k; \vartheta) \tilde{D}_{\gamma \delta}(q; \vartheta) \nonumber \\
& \hspace{1.2cm} \times \bra{0} \phi_{k, \alpha}^{\dagger} \phi_{k,\beta} e^{-i\mathcal{H}t} \phi_{q,\gamma}^{\dagger} \phi_{q,\delta} \ket{0}.
\end{align}
For the finite frequency response only the connected diagrams contribute, which are related with processes involving two excitations that can propagate in time. Analytical results for $R(\omega)$ at different values of the transverse field $h_{y}$ are shown in \figcs{fig:RamanSpectra}{b}{d}.
\begin{figure*}[t]
    \centering
    \includegraphics[width=\textwidth]{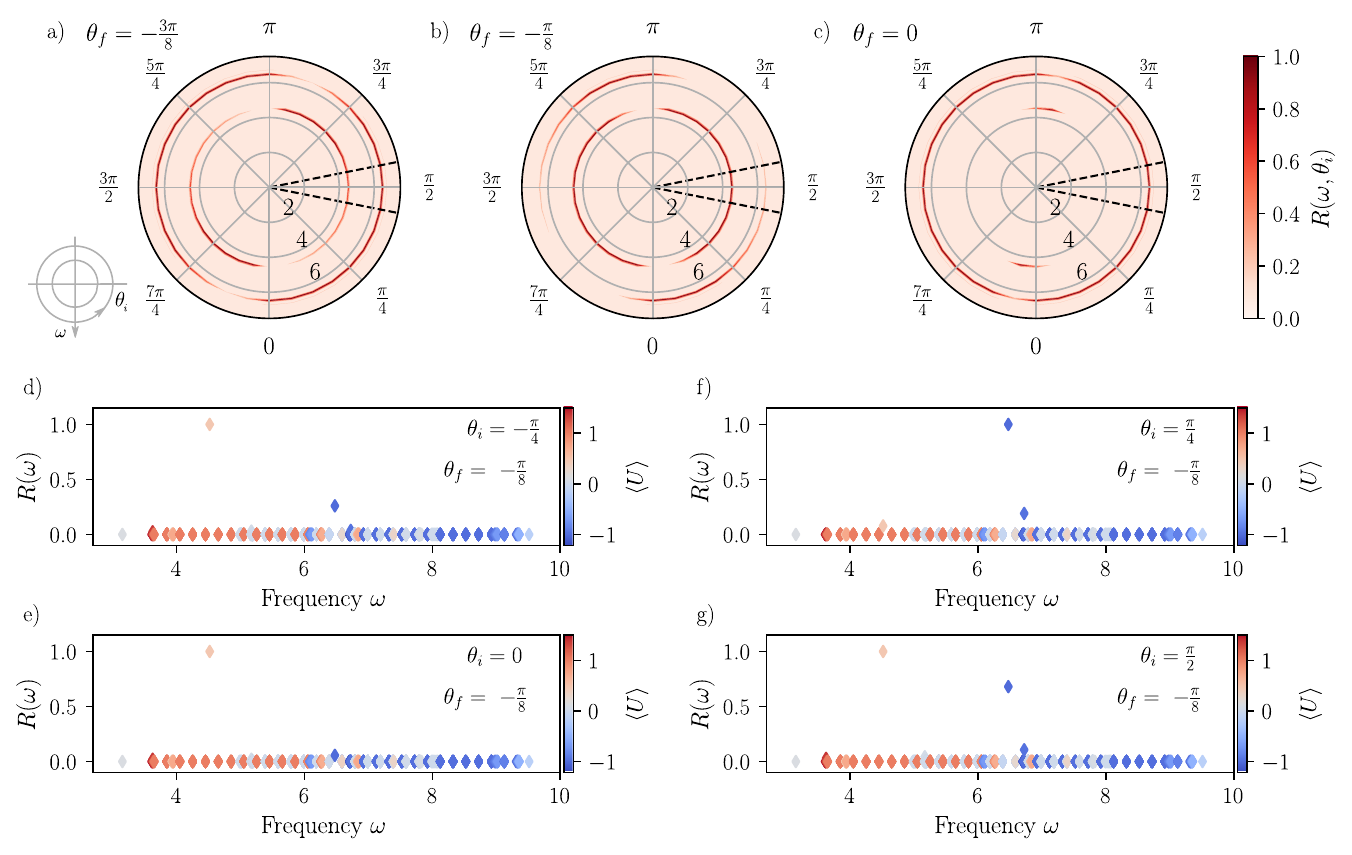}
    \caption{\textbf{Symmetry selection of peaks}  (a) - (c)  Raman response obtained from a few domain wall ansatz as a function of ingoing photon polarization $\theta_{i}$ for outgoing polarization angles $\theta_{f}\in\{-\frac{3\pi}{8}, -\frac{\pi}{8}, 0\}$ and system parameters $\vartheta=\frac{\pi}{8}, (h_{x}, h_{y}, h_{z}) = (0.1, 0., 2 \sin(\vartheta))$ showing the different selection rules for high and low energy bound states.  (d) - (g) The differences between both sets of excitations can be understood from symmetry properties of the eigenstates under pairwise exchange of even and odd bonds $U$. We find eigenstates even under application of $U$ to determine the low energy response (red markers) while the high energy spectrum is governed by odd states (blue markers). Adjusting the photon polarization, e.g. of ingoing photons, enables us to selectively excite even (e) or odd (f) states only or obtain a non-symmetry resolved response (d, g).}
    \label{fig:PolarizationE8}
\end{figure*}\\
From the analytical result we can, moreover, infer information about the angular dependence of the Raman spectra on the polarization of incoming and outgoing photons. Analytical results for the angle-resolved Raman response for different points in the integrable region of the phase diagram are shown in \fig{Fig8:PolarizationDependenceUnconfined}. All spectra correspond to a fixed polarization of incoming photons aligned with the $\hat{z}$-direction of the chain, while tuning the polarization of outgoing photons in the plane spanned by the unit vectors of chain ($\phi_{f}=0$).
The results indicate that obtaining a non-trivial angular dependence for the response requires a finite tilting of the chain, i.e. $\vartheta \neq 0$, as well as finite transverse field components $h_{y}$. The first condition thereby allows to obtain two distinct projectors for even and odd bonds, whose relative values can be tuned by photon polarization. Consequences of the second requirement are more subtle, similar to the effect of magnetic field components $h_{z}$ discussed in \App{sec:PolarizationRaman} finite components $h_{y}$ will cause a staggered character of soliton hoppings allowing for non-trivial symmetries of the bound states at the level of the lattice unit cell.

\section{Polarization Dependence of Raman Responses}
\label{sec:PolarizationRaman}

Having obtained the general expression for the Raman operator, \eq{Eq:RamanDensity}, we now discuss some situations, where the polarization of incoming and outgoing photons can indeed influence the measured response in a nontrivial way. A detailed study of the Raman spectra obtained from the two-domain wall state as a function of the scattered light polarizations are shown in \figcs{fig:PolarizationE8}{a}{c}. We find that the two-domain wall approximation is  able to qualitatively reproduce the iMPS results of \figc{fig3:EnrichedDynamics}{c} of the main text. Moreover, we will now discuss the internal structure of the bound states by constructing an effective model for the domain wall dynamics, following Ref.~\cite{Woodland2023}. 
To this end, we label the two distinct types of domain walls appearing in our system as $\ket{j}_{L} = \ket{\dots \uparrow \uparrow_{j} \downarrow_{j+1} \downarrow \dots}$ and $\ket{j}_{R} = \ket{\dots \downarrow \downarrow_{j} \uparrow_{j+1} \uparrow \dots}$. The dynamics governed by Hamiltonian~(1) for $h_x=0$ within a fixed domain wall sector is given by
\begin{align}
\mathcal{H} \ket{j}_{L} \mapsto &- \sin^{2}\Big(\frac{\vartheta}{2}\Big)\Big[\ket{j-2}_{L} + \ket{j+2}_{L}\Big] \nonumber\\
&+ \Big[\Delta_{L} \sin(\vartheta) - i \frac{h_{y}}{2} - \frac{h_{z}}{2}\Big] \ket{j+1}_{L} \nonumber \\
&+ \Big[-\Delta_{L} \sin(\vartheta) + i \frac{h_{y}}{2} - \frac{h_{z}}{2}\Big] \ket{j-1}_{L}     
\label{eq:DWHopping}
\end{align}
where we introduced $\Delta_{L} = \bra{j}_{L}\sigma_{j-1}^{x}-\sigma_{j+1}^{x}\ket{j}_{L} / 2 = 1$. A similar expression can be obtained for $\ket{j}_{R}$ with $\Delta_{R}=-1$, instead. In the following, we will consider small values of $\vartheta\ll 1$ allowing us to neglect the next-nearest neighbor terms contained in \eq{eq:DWHopping} at leading order. Moreover, we consider only in-plane magnetic fields of our quasi one-dimensional system, i.e. set $h_{y}=0$. When selecting the specific field of $h_{z}=2\sin(\vartheta)$,  the $L$ soliton is only allowed to move within a single unit cell $p$ with bonds $(2p-1, 2p)$ and the $R$ soliton can only hop between neighboring unit cells connected by bonds $(2p, 2p+1)$. The corresponding eigenstates for the single domain wall problem are given by plane wave states built from (anti-)symmetric superpositions in the disconnected two-bond systems
\begin{align}
    \ket{p}_{L}^{\pm} &= \frac{1}{\sqrt{2}} \big[\ket{2p-1}_{L} \pm \ket{2p}_{L} \big] \nonumber\\
     \ket{p}_{R}^{\pm} &= \frac{1}{\sqrt{2}} \big[\ket{2p}_{R} \pm \ket{2p+1}_{R} \big].
\end{align}
Solutions for the two-domain wall problem can be constructed as bound states of two domain walls with given distance
\begin{align}
    \ket{n,k}^{\pm} &= \frac{1}{\sqrt{N}} \sum_{p} e^{i\frac{k}{2} (2p + n)} \ket{p}_{L}^{\pm} \otimes \ket{p+ n}_{R}^{\pm} \nonumber \\ 
    \ket{n,k}^{0}_{\pm} &= \frac{1}{\sqrt{2N}} \sum_{p} e^{i\frac{k}{2} (2p + n)} \Big[ \ket{p}_{L}^{+} \otimes \ket{p+ n}_{R}^{-} \nonumber \\
     & \qquad \pm \ket{p}_{L}^{-} \otimes \ket{p+ n}_{R}^{+} \Big] \quad \text{with} \quad n \ge 1, 
\end{align}
where $N$ denotes the number of unit cells in the system. The associated energies are given by $\omega_{\pm} = 4\cos^{2}(\frac{\vartheta}{2}) \mp h_{z}$ and $\omega_{0} = 4\cos^{2}(\frac{\vartheta}{2})$. With that one can understand the trends of the bound state energies as a function of $h_z$ observed in the spectra of \figcc{fig3:EnrichedDynamics}{a}{b}; see also Ref.~\cite{Woodland2023}. For $n=0$, corresponding to neighboring solitons, we have to additionally take into account a hardcore constraint prohibiting the solitons to occupy the same bond. As a result the eigenstates get altered 
\begin{figure*}
    \centering
    \includegraphics[width=\textwidth]{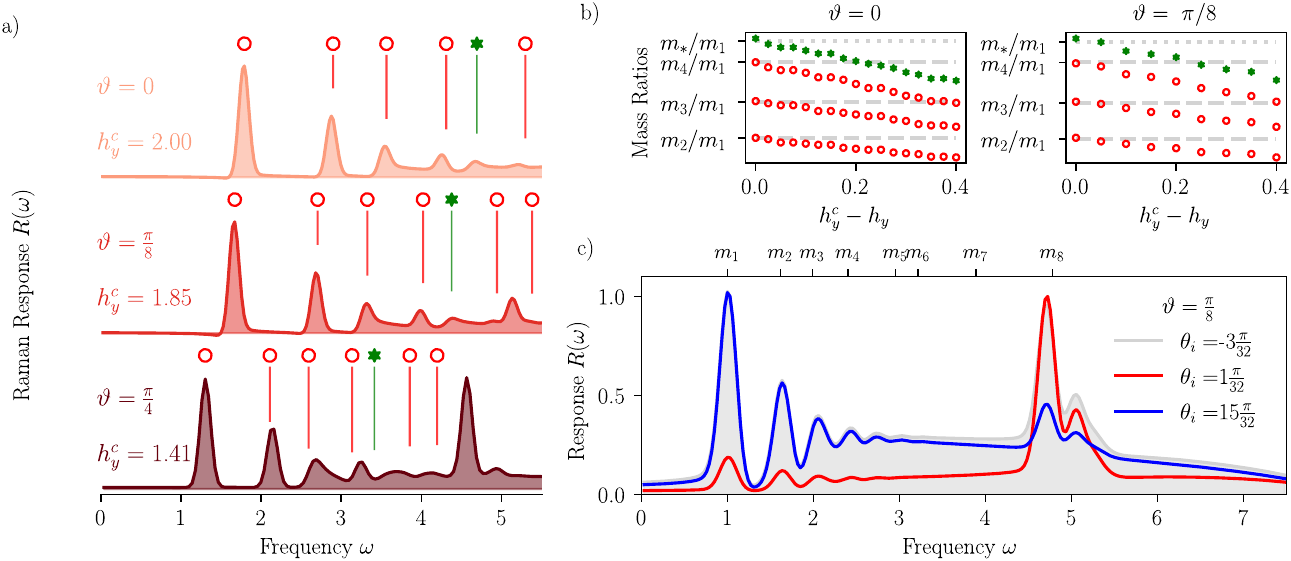}
     \caption{\textbf{Emergent $E_{8}$ field theory at Ising quantum criticality.} (a) Along the Ising quantum critical line the peaks in the Raman response follow the predictions of an anomalous $E_{8}$ field theory (red circles and green stars). A second class of peaks deviating from the predicted mass ratios is resolved at high energies, in particular at larger values of the zigzag angle $\vartheta$. (b) We illustrate the flow of the bound state masses $(m_{2}, m_{3}, m_{4})/m_1$ (red circles) and the summed mass $m_{*}/m_1=(m_{2} + m_{1})/m_1$ (green stars) extracted from Raman spectra. The mass ratios converge to the predictions of the $E_{8}$ field theory as the transverse field $h_{y}$ is tuned to quantum criticality $h_y^c$. (c) Adjusting the photon polarization allows one to modify the relative spectral weight of contributions arising from the Ising field theory and beyond. We tune the polarization such that the Ising term is dominant (blue), the beyond-Ising terms dominate (red) and consider a generic polarization (gray). Selectively addressing higher-energetic contributions can serve as a diagnostic tool to systematically study beyond Ising contributions to the microscopic Hamiltonian. All data has been obtained for the $h_z=0$ and confining fields (a)-(b) $h_{x}=0.25$, (c)  $h_{x}=0.1$. The photon polarization is chosen to be in-plane ($\phi_{i}=\phi_{f}=0$) and characterized by (a) - (b) $(\theta_{i},\theta_{f})=(-\frac{\pi}{2},-\frac{\pi}{4})$ and (c) $\theta_{f}=-\frac{3\pi}{8}$.} 
    \label{figS2:E8Symmetry}
\end{figure*}

\begin{widetext}
\vspace{-\baselineskip}
\begin{align}
    \ket{\epsilon,k}^{\pm} &= \frac{1}{\sqrt{N}} \sum_{p} e^{ikp} \Big[\ket{2p-1}_{L} \otimes \ket{2p}_{R} + \ket{2p}_{L} \otimes \ket{2p+1}_{R} \pm \sqrt{2} \ket{2p-1}_{L} \otimes \ket{2p+1}_{R} \Big] \nonumber \\ 
    \ket{\epsilon, k}^{0} &=  \frac{1}{\sqrt{2N}} \sum_{p} e^{ikp} \Big[\ket{2p-1}_{L} \otimes \ket{2p}_{R} - \ket{2p}_{L} \otimes \ket{2p+1}_{R} \Big]
\end{align}
\end{widetext}
with corresponding eigenenergies of $\epsilon_{\pm}=4\cos^{2}(\frac{\vartheta}{2}) \mp \sqrt{2}h_{z}$ and $\epsilon_{0}=4\cos^{2}(\frac{\vartheta}{2})$. It is worth noting that the large degeneracy in eigenenergies found here, will be lifted by finite longitudinal fields $h_{x}$ penalizing large distances between the domain walls.

We have now understood the flow of eigenenergies as a function of the field $h_z$. As a next step we want to analyze their internal wavefunction. To this end, we consider the pairwise exchange of even and odd bonds
\begin{align}
    U = \sum_{p} \Big[\ket{2p-1}_{L}\bra{2p}_{L} + \ket{2p-1}_{R}\bra{2p}_{R} + \text{h.c.}\Big].
\end{align}
One finds, that $\ket{n,k}^{+}$ and $\ket{\epsilon,k}^{+}$ have positive, $\ket{n,k}^{-}$ and $\ket{\epsilon,k}^{-}$ negative, and $\ket{n,k}_{0}^{\pm}$, $\ket{\epsilon,k}^{0}$ have vanishing expectation values of $U$. In our few-soliton approach we can test the predicted symmetry classification of high and low energy peaks for the eigenstates; \figcs{fig:PolarizationE8}{d}{g}.  For different $\theta_{i}$ values we can selectively excite even (e) or odd (f) eigenstates or both (d) and (g). We indeed resolve a splitting of the measured Raman spectrum into even low-energy and odd high-energy modes. 

\section{Emergent E$_{8}$ Structure at Quantum Criticality}
\label{sec:E8Structure}
\begin{figure*}
    \centering
    \includegraphics[width=\textwidth]{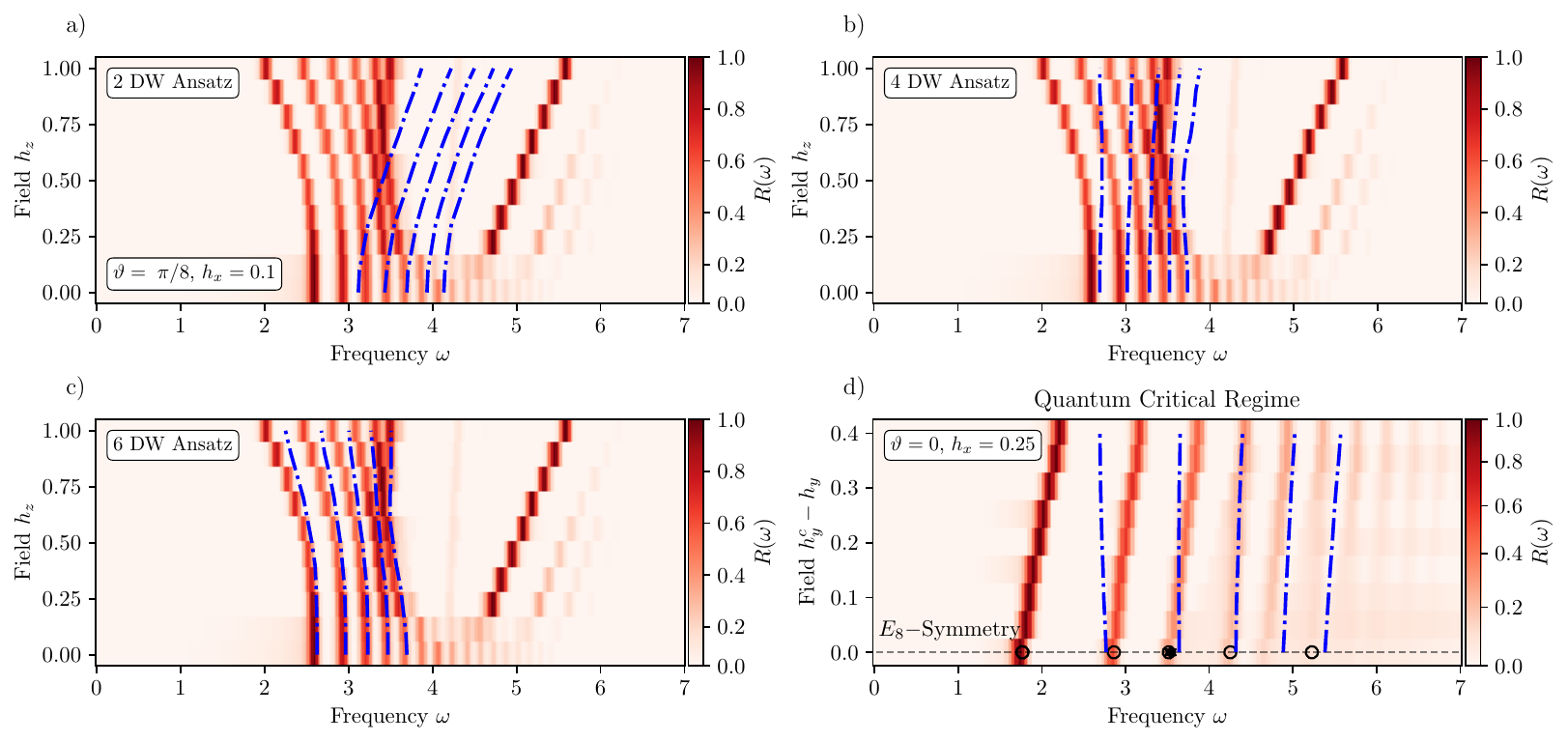}
    \caption{\textbf{Breakdown of the few-soliton ansatz near quantum criticality.} We compare numerically exact iMPS simulations (color plot) and the few-soliton ansatz (blue dash-dotted line) with up to (a) 2, (b) 4 and (c) 6 domain walls for a system with 32 sites. We show the spectrum for $\vartheta=\frac{\pi}{8}$, $h_{x}=0.1, h_{y}=0.5$ and tune $h_{z}$ field. In contrast to the results of \figc{fig1:PhaseDiagram}{c}, which have been obtained in the perturbative regime, we find  that the spectra near criticality deviate from the few-soliton ansatz, even for the largest number of domain walls that we consider, indicating that the bound states possess a more complex structure. (d) In the vicinity of the critical point in the pure Ising case ($\vartheta=0, h_{z}=0, h_{x}=0.25$) all domain wall sectors are expected to contribute to the response, inevitably leading to a breakdown of the few domain-wall ansatz as illustrated for the 6 domain wall case (blue dash-dotted line).}
    \label{fig:BreakdownFK}
\end{figure*}

 As a next step we investigate the structure of the Raman response at quantum criticality. The field theory of the Ising quantum critical point in presence of a weak confinement potential features quasiparticle excitations whose energy resembles the root structure of the anomalous $E_{8}$ Lie algebra~\cite{Rutkevich2010, ZAMOLODCHIKOV1989}. However, even away from the pristine Ising limit, it has been demonstrated both numerically as well as experimentally that the effective description in terms of an $E_{8}$ field theory prevails at low energies, which is an indicator of the underlying Ising quantum critical point~\cite{Kjaell2011, Coldea2011, Amelin2020, Zou2021}. In order to estimate to which extent this signature of quantum criticality can be observed using Raman spectroscopy, we compute the response for $\vartheta \in\{0, \frac{\pi}{8}, \frac{\pi}{4}\}$ at the corresponding critical values of the transverse field $h_{y}^c$ extracted from the analytical solution in the absence of other fields. Results corresponding to a weak longitudinal field of $h_{x}=0.25$ are shown in \figc{figS2:E8Symmetry}{a}. We compare the results for the Raman response with the mass ratios predicted by the $E_{8}$ root structure for the Ising field theory, given by $(m_{2}, m_{3}, m_{4}, m_{5}, m_{6}) = (1.618, 1.989, 2.405, 2.956, 3.218)m_1$~\cite{ZAMOLODCHIKOV1989}. We observe a well-separated peak structure with excellent agreement close to the Ising point for both zigzag angles $\vartheta=\{0, \frac{\pi}{8}\}$. For larger angles we, however, find deviations from the predicted structure, in particular at high energies. This is expected as perturbations to the bare Ising theory will generically produce subleading corrections that determine the energy regime in which the effective field theory is applicable.
 
 To demonstrate that the observed scaling in the spectrum is indeed unique to the critical Ising theory we investigate the flow of the mass ratios (red markers) as well as the collective excitation $m_{*}=m_{1}+m_{2}$ (green marker) extracted from the Raman response for different transverse fields $h_{y}$, both at $\vartheta=0$ and $\vartheta=\frac{\pi}{8}$; \figc{figS2:E8Symmetry}{b}. The $E_{8}$ mass ratios are only reached when $h_y$ approaches $h_y^c$, and are thus a signature of the critical point.
 
 So far we have only considered a fixed scattering geometry by making a single choice for $\theta_{i}, \theta_{f}$. Similar to the advantage in state selection rules given by the external control over photon polarization discussed in the main text, we want to investigate the role of photon polarization in the critical parameter regime of the model. Following the results of \figc{figS2:E8Symmetry}{c} we generically find two types of excitations in the response spectrum. One class consists of excitations following the E$_{8}$ scaling prediction, associated to Ising contributions to the microscopic Hamiltonian. The second class violates the predicted scaling and can therefore be attributed to non-Ising contributions. As demonstrated in \figc{figS2:E8Symmetry}{c} a suitable choice of photon polarization allows one to reduce the intensity of either of the classes in the signal (red vs. blue curve) compared to a generic measurement outcome (gray curve). From this one can systematically study contributions to the microscopic Hamiltonian beyond the dominant Ising theory by enhancing contributions beyond the E$_{8}$-field theory in the response.
 
\section{Complexity of Excitations Near Quantum Criticality}
\label{sec:ExcitationsCriticality}

Here, we study the validity regime of the few domain wall ansatz. The comparison between tensor network results and few-soliton ansatz in the subspace of two and four solitons is discussed in the main text for the perturbative regime; \figc{fig1:PhaseDiagram}{c}. This simple few domain-wall ansatz is only valid deep in the ferromagnetic phase where the excitation gap for domain walls is the dominant scale, i.e., $\vartheta\to 0$ and $h_{y}\ll h_{y}^{c}$. Considering, in contrast parameters closer to the critical regime as shown in \figcs{fig:BreakdownFK}{a}{c} we find that the two domain-wall ansatz cannot reproduce the qualitative trends obtained from tensor network simulations. We investigate the flow of the peak structure as we tune the transverse field $h_{z}$ for values of $\vartheta=\frac{\pi}{8}, h_{y}=0.5$ and $h_{x}=0.1$. Besides a disagreement in energy we  find  increasing bound state masses for both the two and four domain-wall ansatz (blue dash-dotted lines) as $h_{z}$ is increased, whereas iMPS shows the opposite trend. Considering an ansatz involving 6 domain walls captures the right trend, however energies still disagree in particular for large fields $h_z$.  This indicates a more complicated structure of excitations going beyond the simple two soliton bound states. Following up on this prediction we benchmark the Raman response of 6 domain wall ansatz at quantum criticality of the Ising theory ($\vartheta=0, h_{z}=0, h_{x}=0.25$); \figc{fig:BreakdownFK}{d}. When tuning the transverse field $h_{y}$  to the quantum critical point $h_{y}^{c}=2.0$, clear deviations from the predicted $E_{8}$-symmetry are found (black markers). By contrast, iMPS correctly reproduces the scaling of the field theory. This again shows that the confined states possess a complex structure at the quantum critical point.

\end{appendix}

% Hide the References in Table of Contents
\let\oldaddcontentsline\addcontentsline% Store \addcontentsline
\renewcommand{\addcontentsline}[3]{}% Make \addcontentsline a no-op
\bibliography{RamanSpectroscopy}
\let\addcontentsline\oldaddcontentsline% Restore \addcontentsline
%this end  

\end{document}